\documentclass[aip,apl,reprint]{revtex4-2}

\usepackage{amsmath,amsfonts,amssymb}
\usepackage{array,dcolumn}
\usepackage{graphicx}
\usepackage{hyperref}
\usepackage{upgreek}

\begin{document}

\title{The effects of ion implantation damage to photonic crystal optomechanical resonators in silicon}

\author{Cliona Shakespeare}
\affiliation{Department of Physics and Nanoscience Center, University of Jyv{\"a}skyl{\"a}, P.O. Box 35, FI-40014 University of Jyv{\"a}skyl{\"a}, Finland}

\author{Teemu Loippo}
\affiliation{Department of Physics and Nanoscience Center, University of Jyv{\"a}skyl{\"a}, P.O. Box 35, FI-40014 University of Jyv{\"a}skyl{\"a}, Finland}

\author{Henri Lyyra}
\affiliation{Department of Physics and Nanoscience Center, University of Jyv{\"a}skyl{\"a}, P.O. Box 35, FI-40014 University of Jyv{\"a}skyl{\"a}, Finland}

\author{Juha T. Muhonen}
\email{juha.t.muhonen@jyu.fi}
\affiliation{Department of Physics and Nanoscience Center, University of Jyv{\"a}skyl{\"a}, P.O. Box 35, FI-40014 University of Jyv{\"a}skyl{\"a}, Finland}

\date{\today}

\begin{abstract}
Optomechanical resonators were fabricated on a silicon-on-insulator (SOI) substrate that had been implanted with phosphorus donors. The resonators' mechanical and optical properties were then measured (at 6 kelvin and room temperature) before and after the substrate was annealed.
All measured resonators survived the annealing and their mechanical linewidths decreased while their optical and mechanical frequencies increased. This is consistent with crystal lattice damage from the ion implantation causing the optical and mechanical properties to degrade and then subsequently being repaired by the annealing. We explain these effects qualitatively with changes in the silicon crystal lattice structure. We also report on some unexplained features in the pre-anneal samples.
In addition, we report partial fabrication of optomechanical resonators with neon ion milling.
\end{abstract}

\maketitle

Ion implantation is a widely-used technology for doping of semiconductors with applications varying from microchips \cite{Poate2002,rubin_ion_nodate}, to solar panels \cite{rohatgi_high-throughput_2012}, to quantum computing \cite{Donkelaar2015}. As ion implantation is a somewhat destructive method, basically bombarding the substrate with high-energy ions, the process inevitably damages the crystal structure of the substrate. This problem is especially acute with silicon, as it is easily damaged and can become amorphous \cite{williams_ion_1998}.

The implantation damage is usually healed with a post-implantation anneal \cite{ziegler_high_1985} that both allows the crystal to repair itself and activates any substitutional donor atoms as they can then take their place in the crystal structure. Here we study the effects of the implantation damage and subsequent annealing to the mechanical and optical properties of suspended photonic crystal nanobeams.

The degradation of the optical properties of silicon from ion implantation is of importance to any devices combining ion implantation and photonics in silicon, a field that is growing in importance as single-photon emitters in silicon are being actively investigated \cite{Chen2020,Bergeron2020,Weiss2021,Durand2021}, and could also affect other silicon photonics devices such as vertical-cavity surface-emitting lasers \cite{siriani_phase_2010}. Similarly, the degradation of mechanical properties affects any studies combining implanted ions and mechanical systems (such as hybrid structures combining donor spin qubits \cite{Pla2012,Muhonen2014} with phonon buses \cite{Patel2018}).

\begin{figure}
    \centering
    \includegraphics[width=0.48\textwidth]{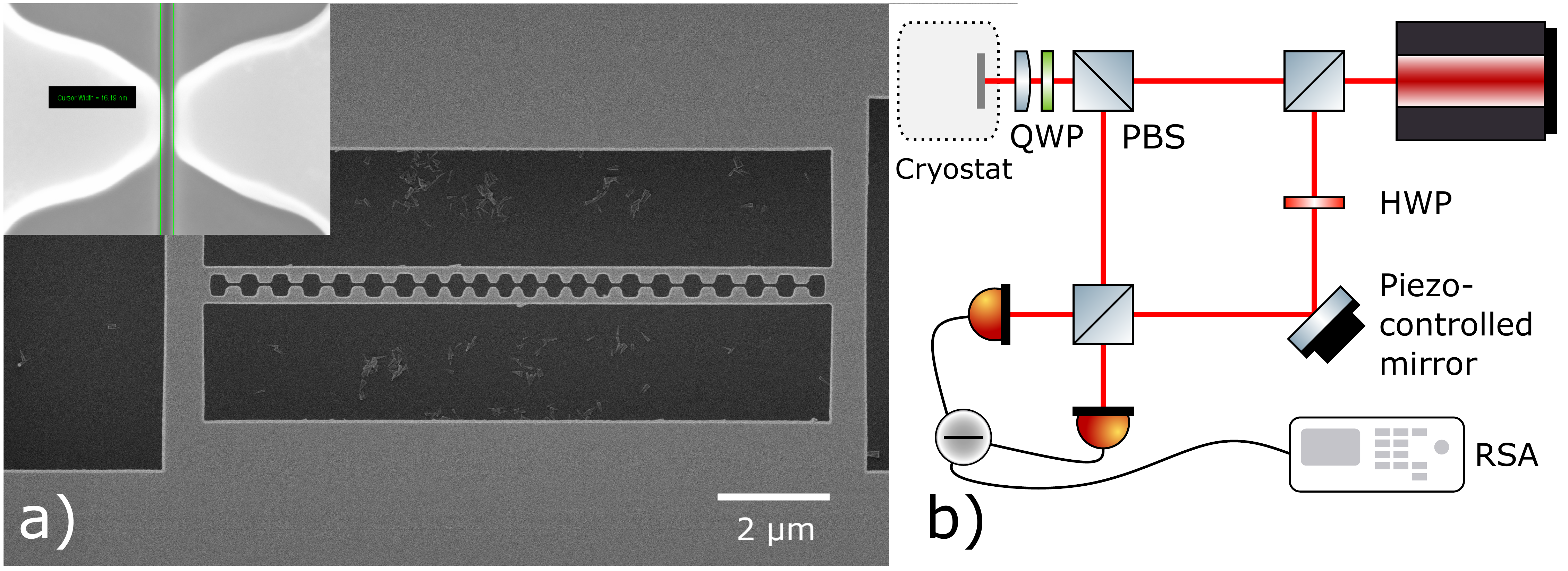}
    \caption{a) Scanning electron microscope (SEM) image of an example nanobeam. Gray parts are free hanging silicon and the dark parts show the substrate 3~µm below. The optical cavity is defined in the center of the silicon nanobeam where the photonic crystal lattice constant is smaller. The two beam halves have been separated by a small gap. Inset: zoom to the gap between nanobeam teeth showing the roughly 20~nm separation before suspension that was defined with neon ion milling. b) Schematic of the interferometer used for measurements; QWP refers to a quarter-wave-plate, HWP to a half-wave-plate, PBS to a polarizing beam splitter and RSA to a realtime spectrum analyzer.}
    \label{fig:beam}
\end{figure}

The resonators we study, as shown in Fig. \ref{fig:beam}, are sliced photonic crystal nanobeams, the design of which was presented in Ref. \onlinecite{leijssen_strong_2015}. Similar structures were also used in later studies \cite{Leijssen2017,Muhonen2019}. The photonic crystal in the center of the nanobeam supports different optical modes than the photonic crystal on the sides which then work as Bragg reflectors for a targeted cavity mode. Note that here we do not have a tapering region between the mirrors and the cavity, although it has been shown to decrease the optical linewidth considerably \cite{Leijssen2017}. The cavity optical mode has an electric field maximum at the center of the silicon beam, which is also the location of the highest mechanical displacement (for the fundamental mechanical mode). The small vacuum gap between the two beam halves leads to strong field confinement in the gap, and a large frequency dependence of the optical mode on the gap size. This gives strong optomechanical coupling for the mechanical mode where the two beam halves move asymmetrically in-plane (breathing mode).

The samples were fabricated on a Silicon-On-Insulator (SOI) wafer with 220~nm device layer and 3~µm buried oxide. The device layer was implanted with phosphorus ions. The ion implantation \cite{Surrey} aimed for a constant donor density of $10^{17}$~ions/cm$^3$ in the device layer and was done in three steps: 110~keV energy with a dose of $1.5 \times 10^{12}$~ions/cm$^2$, 40~keV with $5 \times 10^{11}$~ions/cm$^2$ and 20~keV with $1.5 \times 10^{11}$~ions/cm$^2$. The wafer was not annealed after implantation, leaving any potential lattice damage intact.

The wafer was then diced and optomechanical resonators were fabricated on chips. The chips were spin coated first with Microchem 950~k PMMA A4 resist, which was then baked for 3~min on a hotplate at 160$^\circ$C, then with Allresist AR-PC 5090.02 (Electra 92), which was baked for 1~min at 80$^\circ$C on a hotplate. The resist was patterned with a Raith eLine electron beam lithography tool. The PMMA was developed for 20~s in a 2:1 mixture of IPA and MIBK. Before development, the Electra 92 resist was washed off. The etching was done with an Oxford Plasmalab 80 reactive ion etcher with 80~sccm SF$_6$ and 20~sccm O$_2$ at 30~mTorr at $-$100$^\circ$C using liquid nitrogen cooling. 

The gap between the beam halves was afterwards milled through with 25~kV neon ions in a Zeiss Orion NanoFab helium ion microscope (HIM). Simulations of the neon milling with SRIM \cite{srim} (28~keV ion energy, 3-4~pA ion current, 7.5~nC/µm$^2$ area dose) show a lateral ion path of up to 30~nm, so there should be only a moderate amount of neon ions in the silicon nanobeams from the milling. Finally the optomechanical structures were released by wet etching the oxide layer with 48~\% HF for 4~min. An example of a released structure is shown in Fig. \ref{fig:beam} (a).

A picture of the milled gap between the beam halves is shown in the inset of Fig. \ref{fig:beam} (a) where the high-resolution enabled by neon milling is demonstrated as the gap is only 20 nm wide. This would be beneficial for optomechanical applications as the optomechanical coupling strength in these devices grows with decreasing gap size. However, SEM images taken after the HF release show an opening of roughly 50 nm. This could be due to the beam halves bending outwards from stress or charging but requires further study.

The samples were placed inside a Montana Instruments C2 s50 cryostat and measured with a balanced homodyne interferometer (schematic in Fig. \ref{fig:beam}(b)) using near infrared wavelengths. The optical cavity was coupled into from above with a free space laser beam and the reflected light was interfered with a reference laser beam. The interference signal as a function of laser wavelength was measured with a subtracting detector and a spectrum analyzer. Lorentzian line shapes were fitted to the brightest mechanical peak visible in each spectrum (using the Python package SciPy \cite{scipy}) and the optical and mechanical resonance frequencies and linewidths were extracted from these fits. As we do not lock the phase of the interferometer but average over all the phases, our mechanical peaks split with the frequency of the piezo sweep (c.~100 Hz) and the effective optical linewidth is modified, both of which we account for in our fitting procedures.

After measuring the resonators at both room temperature and 6~K, the samples were taken out and annealed in a tube furnace for 30~min at 600$^\circ$C, 15~s at 1000$^\circ$C, and 15~min at 500$^\circ$C. All annealing was done in an argon atmosphere with 1~atm pressure. The chip was left to cool overnight, after which the chip was loaded into the cryostat and the measurements at room temperature and 6~K repeated.

\begin{figure}
    \centering
    \includegraphics[width=0.48\textwidth]{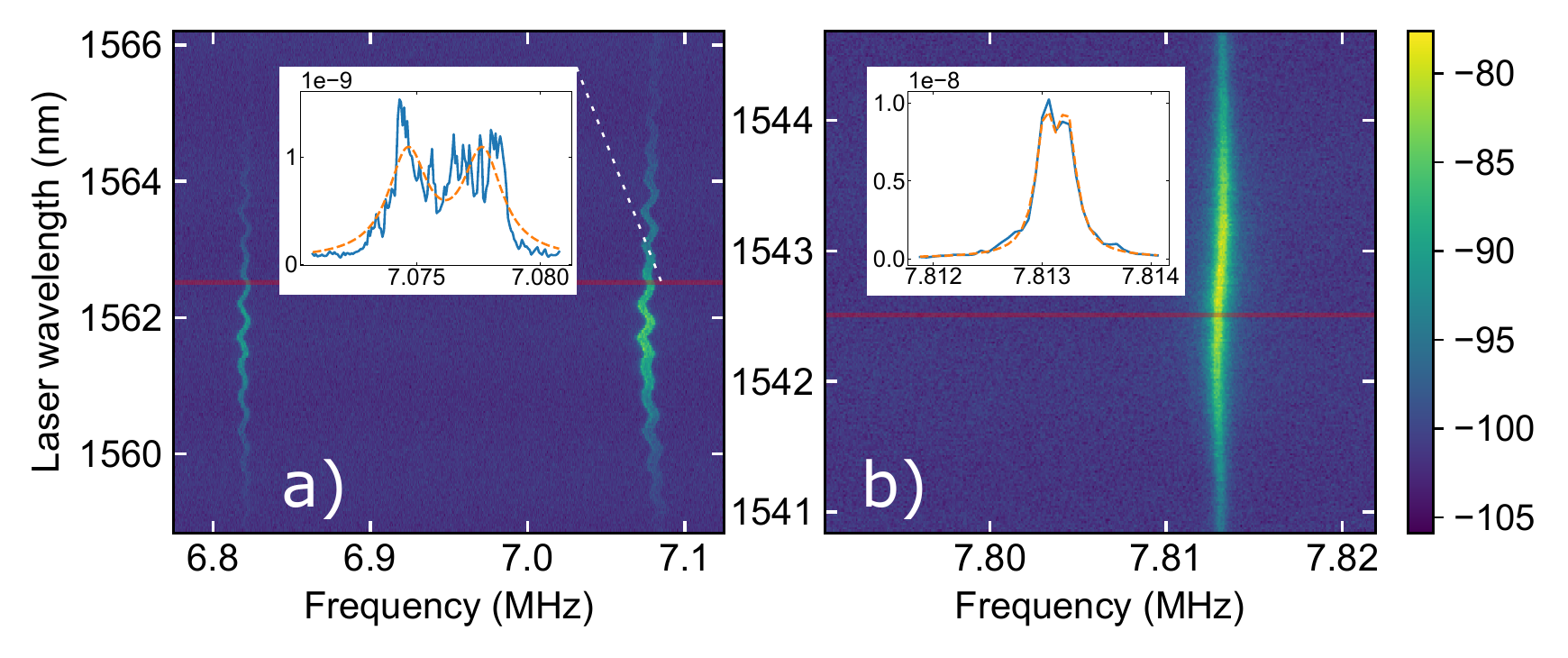}
    \caption{Resonator 2 spectrogram measured at 6~K before (a) and after (b) annealing as a function of the laser wavelength. The colorbar shows the power of the interferometer signal in dB. Crosscuts of the spectra (on a linear scale) are shown in the insets with Lorentzian fits (dashed line). The post-anneal sample exhibits much clearer Lorentzian resonance. The reference levels for the signal are arbitrary.}\label{fig:rez2}
\end{figure}

The wavelength dependent spectrograms of the interferometer signal from one of the resonators are shown in Fig. \ref{fig:rez2}. The mechanical response is visible when the laser wavelength is close to the optical cavity resonance frequency. An unexpected phenomenon was an oscillation of the mechanical resonance frequency as a function of the laser wavelength in the pre-anneal samples at 6~K, as shown in Fig. \ref{fig:rez2}(a). This effect was present in both resonators though much more noticeable with resonator 2 and disappeared after annealing. The oscillation was consistent between consecutive measurements (meaning the same wavelength would produce same response) but varied when the laser spot position (with regards to the silicon nanobeam) was moved. Changing the laser power by a factor of 5 had no significant effect on the phenomenon, implying it is not a thermal effect. We also note that these pre-anneal spectrograms at 6~K seem like they are a combination of two different peaks and we have fitted them accordingly for the analysis below.

\begin{figure}
    \centering
    \includegraphics[width=0.48\textwidth]{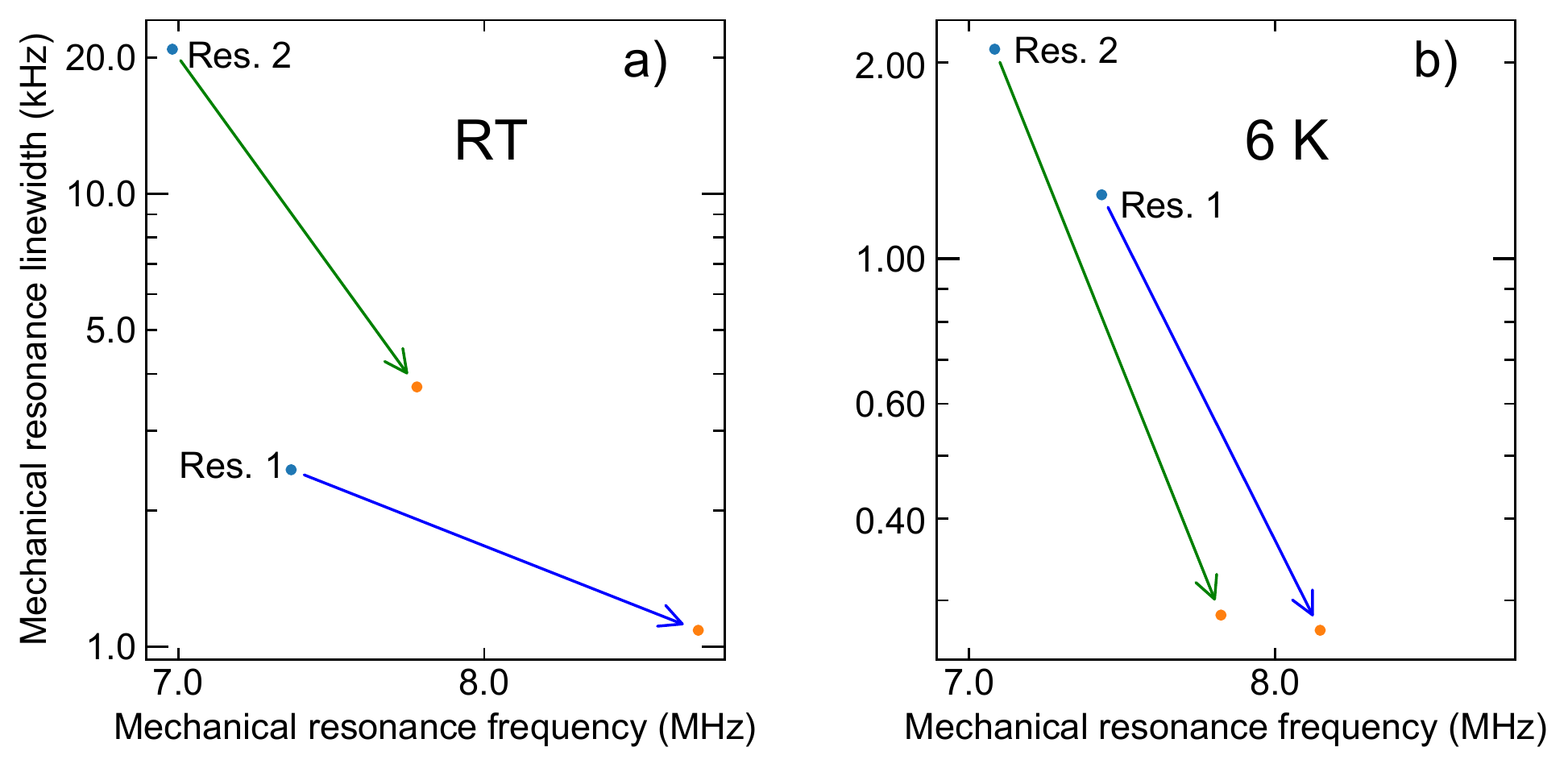}
    \caption{Plots of the resonators' mechanical resonance frequencies and linewidths before and after the annealing at room temperature (a) and 6~K (b). The values before and after annealing have been connected with blue (Resonator 1) and green (Resonator 2) arrows for clarity.}\label{fig:mech}
\end{figure}

After annealing the mechanical resonance frequencies increased consistently by 10-20 \% for both samples at both temperatures, as pictured in Fig. \ref{fig:mech}, with pre-anneal frequencies lying in the range of 7-8 MHz. An increase in resonance frequency implies a stiffer resonator, i.e. a higher Young's modulus. While the Young's modulus for silicon depends on crystal orientation \cite{hopcroft_what_2010}, it is higher for crystalline silicon in the fabrication direction of our device --- [110] --- than it is for polycrystalline silicon, meaning that the shift in resonance frequency can be at least qualitatively explained by the lattice structure being damaged by the ion implantation, making it in effect more polycrystalline-like and then being reconstructed by the annealing.

To study the effect further, we made COMSOL simulations of the beams' mechanical resonance frequencies for both polycrystalline Si and Si with the same crystal direction as our sample using Young's moduli from Ref. \onlinecite{hopcroft_what_2010}. The simulated resonance frequency for polycrystalline silicon nanobeams, 6.7~MHz, was lower than the 6.9~MHz of crystalline silicon, as expected. However, the measured shifts are consistently higher than this effect alone would predict.

The mechanical linewidths for both resonators decreased after the anneal. Interestingly all measured modes (including unshown less bright modes) ended up with a linewidth of $0.2-0.3$~kHz at cryogenic temperatures after the anneal, possibly indicating a limit where some mechanism not related to temperature or crystal damage is determining the linewidth (such as clamping losses). Both linewidths decreased by roughly a factor of 5. In room temperature measurements resonator 1's linewidth improved by a factor of 2 whereas resonator 2 had an improvement by factor of 6. The trend is consistent with the lattice structure healing from implantation damage during the annealing, as lattice defects can cause mechanical damping \cite{shin_mechanical_2015}.

\begin{figure}
   \centering
   \includegraphics[width=0.48\textwidth]{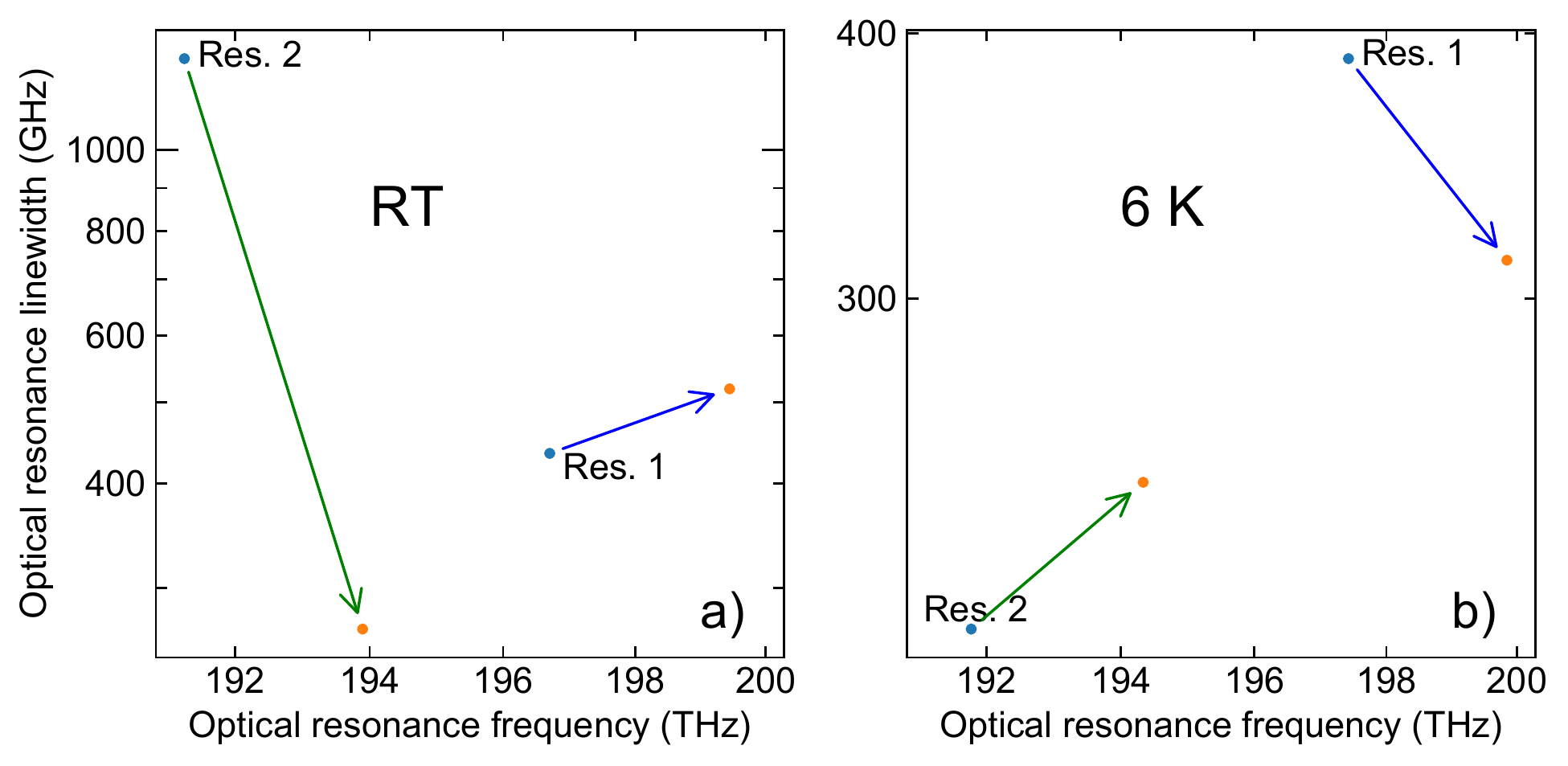}
   \caption{Plots of the resonators' optical resonance frequencies and linewidths before and after the annealing at room temperature (a) and 6~K (b). The values before and after annealing have been connected with blue (Resonator 1) and green (Resonator 2) arrows for clarity.}\label{fig:opt}
\end{figure}

A similar plot for the optical properties is shown in Fig. \ref{fig:opt}. The optical cavity resonance frequency and cavity linewidth can be extracted from the strength of the optomechanical signal. The optical resonance frequencies increased by $\sim$1\% during the anneal. Here there are two effects that might be playing a role: i) a decrease in the effective refractive index or ii) change in the equilibrium gap size between the beams due to mechanical reasons. We unfortunately cannot completely untangle these effects. The refractive index of polycrystalline silicon is higher than that of single-crystal silicon \cite{jones_electrical_1984}. Thus recrystallization of the silicon lattice could lead to a lowering of refractive index and a resonance shift towards higher optical frequencies in accordance with our results. We also imaged the sample with SEM before and after annealing and saw no visible change in the gap size. Note that the doping concentration should not affect the effective refractive index at these wavelengths and doping levels \cite{Auslender2017}.

For the optical linewidth we did not recover any consistent trend. We note that there could be two competing factors here. Silicon crystal lattice damage can be expected to cause photon losses to increase, and polycrystalline silicon also has a higher coefficient of absorption at these wavelengths \cite{harbeke_growth_1984}, which means that annealing should decrease the linewidth. On the other hand, doping activation during annealing would be expected to cause photon losses to increase. We do note that after annealing there seems to be no significant temperature dependence in the optical linewidth.

In conclusion, we have shown that post-implantation annealing has a significant effect on the optical and mechanical properties of silicon photonic crystal nanobeams. The exact mechanisms of the effects we demonstrated still require further study but can qualitatively be explained by implantation damage making the silicon more polycrystalline-like and annealing then recovering the single-crystalline structure. These findings will have relevance to any nanophotonic device that combines ion implantation with photonic crystal structures, particularly the emerging field of spin-photonics in silicon as well as to any structures combining ion implantation and mechanical devices in silicon.
The oscillation of the mechanical resonance frequency as a function of the laser wavelength in the pre-anneal samples is an intriguing phenomenon that could bear further investigation.

\begin{acknowledgements}
We acknowledge Luke Antwis and Roger Webb for help with the ion beam implantations. This project has received funding from the European Research Council (ERC) under the European Union’s Horizon 2020 research and innovation programme (grant agreement No 852428) and by Academy of Finland Grant No 321416.
\end{acknowledgements}

\bibliography{bibliography}

\end{document}